\documentclass[preprint,showpacs,preprintnumbers]{revtex4}
\usepackage{graphicx}
\usepackage{dcolumn}

\begin{document}

\title{Slow Light in Artificial Hybrid Molecules }
\author{Zhien Lu}
\email{zhienlu @sjtu.edu.cn}
\author{Ka-Di Zhu}
\email{zhukadi@sjtu.edu.cn}
\affiliation{Department of Physics, Shanghai Jiao Tong University, 800 DongChuan Road,
Shanghai 200240, China }
\date{\today}

\begin{abstract}
The optical properties of hybrid molecules composed of semiconductor and
metal nanoparticles with a weak probe in a strong pump field are
investigated theoretically. Excitons in such a hybrid molecule demonstrate
novel optical properties due to the coupling between exciton and plasmon. It
is shown that a non-absorption hole induced by coherent population
oscillation appears at the absorption spectrum of the probe field and there
exists slow light effect resulting in the great change of the refractive
index. The numerical results indicate that with the different
center-to-center distance between the two nanopaticles the slow light
effects are greatly modified in terms of exciton-plasmon couplings.
\end{abstract}

\maketitle

% Force line breaks with

% It is always \today, today,
%  but any date may be explicitly specified

% PACS, the Physics and Astronomy
% Classification Scheme.
%\keywords{Suggested keywords}%Use showkeys class option if keyword
%display desired

\section{Introduction}

Due to the rapid advances of nanotechnology, it is possible to combine
nanocrystal of various material with very different characteristics in one
superstructures\cite{1,2,3}, then the investigation on the properties of
artificial hybrid molecules has attracted much interest in recent years\cite%
{4,5,6,7,8}. One of the central problems related to these hybrid complex is
the interaction between the nanoscale building blocks. If individual
nanoparticles do not exchange carriers, then the interparticle Coulomb
interaction becomes the main mechanism of coupling and can strongly change
the physical properties of these nanoscale building blocks. On the other
hand, there is a great deal of researches devoted to investigate the slow
light based on electromagnetically induced transparency (EIT)\cite{9,10,11}
and coherent population oscillation (CPO)\cite{12} in various of potential
materials. During the last decade, coherent population oscillation(CPO)
approach have proven to be a power technique that can eliminate the
absorption at the resonant frequency of transition and dramatically change
the refractive index which leads to the slowdown of light speed in various
of mediums\cite{15,16,17}. One of the more effective thing is that the slow
light based on the CPO approach can be achieved at room temperature,
comparing with that based on electromagnetically induced transparency
method. But most materials mentioned above are only focused on the
conventional materials, such as optical fiber, Pb vapours, room-temperature
solid and low dimensional semiconductors. The investigation of slow light in
hybrid complex is obviously different from that of the conventional low
dimensional semiconductors, then it is hoped that some novel optical
properties will be found in the hybrid complex.

In the present paper, we will study the slow light in hybrid complex
composed of a metal nanoparticle(MNP) and a semiconductor quantum dot(SQD).
Elementary excitations in a SQD and a MNP, an exciton and a plasmon, have
very different properties. The basic excitations in the MNP are surface
plasmons with a continuous spectrum. In SQDs, the excitations are discrete
interband excitons. There is no direct tunneling between the MNP and the
SQD. However, long-range coulomb interaction couples the excitons and
plasmons\cite{18}. The effect of coupling between excitons and plasmons can
strongly alter the optical properties of superstructure compared to a single
SQD.

The structure of the article is as follows: In Sec.II, the theoretical model
is given. The numerical results are presented in Sec.III. In Sec.IV, a
summary of our results is given.

\section{Theory}

We now consider a hybrid molecule compose of a spherical MNP with radius $a$
and a spherical SQD with radius $r$ with a weak probe in a strong pump
field. The center-to-center distance between the two nanoparticles is $R$.
For the SQD, we assume a two-level model for such a SQD which consists of
the ground state$\left\vert 0\right\rangle $ and the first excited
state(single exciton)$\left\vert 1\right\rangle $. The two-level scheme with
coherent drive in atomic systems has been studied by Yu et al\cite{19} which
measured the emission spectrum of two-level-like Ba atoms driven by a
continuous wave bichromatic field. The hybrid molecule via exciton interacts
with a probe field of the frequency $\omega _{s}$ and is coherently driven
by a strong control field of frequency $\omega _{c}$. As usual, this
two-level system can be characterized by the \ pseudospin$-\frac{1}{2}$
operators $S^{\pm }$ and $S^{z}$. Then the Hamiltonian of the system in a
rotating frame at the control field frequency $\omega _{c}$ reads as follows:%
\begin{equation}
H=\hbar (\omega _{ex}-\omega _{c})S^{z}-\hbar \Omega (S^{+}+S^{-})-\mu
(S^{+}E_{s}e^{-i\delta t}+S^{-}E_{s}^{\ast }e^{i\delta t}),
\end{equation}
where $\hbar \omega _{ex}$ is the energy of exciton binding, $\Omega =\mu
E_{SQD}/\hbar $, $\mu $ is the interband dipole matrix element, $E_{SQD}$ is
the total field felt by the SQD and is given by%
\begin{equation}
E_{SQD}=E_{c}+\frac{S_{\alpha }P_{MNP}}{\varepsilon _{eff_{1}}R^{3}}
\end{equation}%
with $\varepsilon _{eff_{1}}=\frac{2\varepsilon _{0}+\varepsilon _{s}}{%
3\varepsilon _{0}}$, $\varepsilon _{0}$ and $\varepsilon _{s}$ are the
dielectric constants of the background medium and SQD, respectively, $E_{c}$
is the slowly varying envelope of the control field and $S_{\alpha }=2$ for
electric field polarizations $\alpha =z$ which direction corresponds to the
axis of the hybrid molecule. The dipole $P_{MNP\text{ }}$ comes from the
charge induced on the surface of the MNP. It depends on the total electric
field with superposition of the control and the dipole field due to the
SQD,and is given by%
\begin{equation}
P_{MNP}=\gamma a^{3}[E_{c}+\frac{S_{\alpha }P_{SQD}}{\varepsilon
_{eff_{2}}R^{3}}],
\end{equation}
where $\gamma=\frac{\varepsilon _{m}(\omega )-\varepsilon _{0}}{3\varepsilon
_{0}+\varepsilon _{m}(\omega )}$, $\varepsilon _{eff_{2}}=\frac{2\varepsilon
_{0}+\varepsilon _{s}}{3}$. $\varepsilon _{m}(\omega )=1-\frac{\omega
_{p}^{2}}{\omega ^{2}}$ is the dielectric constant of the metal, and $\omega
_{p}$ is the surface plasmon frequency of the MNP. The dipole moment of the
SQD is expressed via the off-diagonal elements of the density matrix:$%
P_{SQD}=\mu S^{-}$\cite{22}. $E_{s}$ is the slowly varying envelope of the
probe field. $\delta =\omega _{s}-\omega _{c}$ is the detuning of the probe
and the control field. $\omega _{s}$ is the frequency of the probe field.
The temporal evolutions of the exciton on the SQD are determined by the
Heisenberg equation of motion, and are given by

\begin{equation}
\frac{dS^{-}}{dt}=-i\Delta S^{-}-2i\Omega S^{z}-\frac{2i\mu }{\hbar }%
S^{z}E_{s}e^{-i\delta t},
\end{equation}%
\begin{equation}
\frac{dS^{z}}{dt}=i\Omega (S^{+}-S^{-})+i\mu (S^{+}E_{s}e^{-i\delta
t}-S^{-}E_{s}^{\ast }e^{i\delta t}),
\end{equation}%
where $\Delta =\omega _{ex}-\omega _{c}$. In what follows we ignore the
quantum properties of $S^{-}$, $S^{z}$ \cite{23,24,25}, then the
semiclassical equations read as follows%
\begin{equation}
\frac{dS^{-}}{dt}=(-\frac{1}{T_{2}}-i\Delta )S^{-}-2i\Omega S^{z}-\frac{%
2i\mu }{\hbar }S^{z}E_{s}e^{-i\delta t},
\end{equation}%
\begin{equation}
\frac{dS^{z}}{dt}=\frac{1}{T_{1}}(S^{z}+\frac{1}{2})+i\Omega
(S^{+}-S^{-})+i\mu (S^{+}E_{s}e^{-i\delta t}-S^{-}E_{s}^{\ast }e^{i\delta
t}),
\end{equation}%
where $T_{1}$ is the exciton lifetime, T$_{2}$ is the exciton dephasing
time. For simplicity we define $p=\mu S^{-}$, $w=2S^{z}$, and then we have%
\begin{equation}
\frac{dp}{dt}=-(\frac{1}{T_{2}}+i\Delta )p-\frac{i\mu ^{2}}{\hbar }wE,
\end{equation}%
\begin{equation}
\frac{dw}{dt}=-\frac{1}{T_{1}}(w+1)+4Im(pE^{\ast }),
\end{equation}%
where $E=E_{SQD}+E_{s}e^{-i\delta t}$. In order to solve Eqs.(8) and (9), we
make the ansatz\cite{26}
\begin{equation}
p=p_{0}+p_{1}e^{-i\delta t}+p_{-1}e^{i\delta t},
\end{equation}%
\begin{equation}
w=w_{0}+w_{1}e^{-i\delta t}+w_{-1}e^{i\delta t},
\end{equation}%
on substituting Eqs.(10) and (11) into Eqs.(8) and (9) and on working to the
lowest order in $E_{s}$ but to all orders in $E_{c}$, we obtain in the
steady state:

\begin{equation}
0=-(\frac{1}{T_{2}}+i\Delta )p_{0}-\frac{i\mu ^{2}B}{\hbar }w_{0}p_{0}-\frac{%
i\mu ^{2}A}{\hbar }E_{c},
\end{equation}
\begin{equation}
-i\delta p_{1}=-(\frac{1}{T_{2}}+i\Delta )p_{1}-\frac{i\mu ^{2}B}{\hbar }%
(w_{0}p_{1}+w_{1}p_{0})-\frac{i\mu ^{2}}{\hbar }(Aw_{1}E_{c}+w_{0}E_{s}),
\end{equation}
\begin{equation}
i\delta p_{-1}=-(\frac{1}{T_{2}}+i\Delta )p_{-1}-\frac{i\mu ^{2}B}{\hbar }%
(p_{0}w_{-1}+p_{-1}w_{0})-\frac{i\mu ^{2}A}{\hbar }w_{-1}E_{c},
\end{equation}
\begin{equation}
0=-\frac{1}{T_{1}}(w_{0}+1)+\frac{2i}{\hbar }[A(p_{0}^{\ast
}E_{c}-p_{0}E_{c}^{\ast })],
\end{equation}%
\begin{equation}
-i\delta w_{1}=-\frac{1}{T_{1}}w_{1}+\frac{2i}{\hbar }[A(p_{-1}^{\ast
}E_{c}-p_{1}E_{c}^{\ast })+B(p_{-1}^{\ast }p_{0}-p_{1}p_{0}^{\ast
})+p_{0}^{\ast }E_{s}],
\end{equation}
\begin{equation}
i\delta w_{-1}=-\frac{1}{T_{1}}w_{-1}+\frac{2i}{\hbar }[A(p_{1}^{\ast
}E_{c}-p_{-1}E_{c}^{\ast })+B(p_{1}^{\ast }p_{0}-p_{-1}p_{0}^{\ast
})-p_{0}E_{s}^{\ast }],
\end{equation}
where $A=1+\frac{\gamma a^{3}S_{\alpha }}{\varepsilon _{eff_{1}}R^{3}}$, $B=%
\frac{\gamma a^{3}S_{\alpha }^{2}}{\varepsilon _{eff_{1}}\varepsilon
_{eff_{2}}R^{6}}$.

From the solution of Eqs.(12)-(17), we can yield
\begin{equation}
w_{1}=-\frac{2\frac{\mu ^{2}}{\hbar ^{2}}Aw_{0}E_{s}E_{c}^{\ast }}{D(\delta
_{c})}T_{2}^{2}[2-i(\delta _{c}+B_{c}w_{0})][1+i(\Delta
_{c}+B_{c}w_{0})][(1-i(\delta _{c}+\Delta _{c}+B_{c}w_{0})],
\end{equation}%
\begin{eqnarray}
p_{1} &=&\frac{\mu ^{2}E_{s}T_{2}}{\hbar \lbrack 1+i(\Delta _{c}-\delta
_{c}+B_{c}w_{0})]}\times \{-iw_{0}+\frac{2A^{2}\Omega _{c}^{2}w_{0}}{%
D(\delta _{c})}(1+i\Delta _{c})[1-i(\Delta _{c}+\delta _{c}+B_{c}w_{0})] \\
&&\times \lbrack 2-i(\delta _{c}+B_{c}w_{0})]\},
\end{eqnarray}%
where%
\begin{eqnarray*}
D(\delta _{c}) &=&(2-i\delta _{c})[1+(\Delta
_{c}+B_{c}w_{0})^{2}][(1-i\delta _{c})^{2}+(\Delta _{c}+B_{c}w_{0})^{2}] \\
&&+4A^{2}\Omega _{c}^{2}(1-i\delta _{c})(1+\Delta _{c})^{2},
\end{eqnarray*}%
where $T_{1}=\frac{T_{2}}{2}$, $\delta _{c}=\delta T_{2}$, $\Omega
_{c}^{2}=\mu ^{2}\frac{\left\vert E_{c}\right\vert ^{2}}{\hbar ^{2}}%
T_{2}^{2} $, $\Delta _{c}=\Delta T_{2}$, $B_{c}=\frac{\mu ^{2}B}{\hbar }%
T_{2} $. The linear optical susceptibility can be expressed by
\begin{equation}
\chi _{eff}^{(1)}=N\frac{p_{1}}{E_{s}}=\frac{N\mu ^{2}T_{2}}{\hbar }\chi
^{(1)}(\omega _{s}),
\end{equation}%
where $N$ is the number density of hybrid molecule and the dimensionless
susceptibility is given by

\begin{eqnarray}
\chi ^{(1)}(\omega _{s}) &=&\frac{1}{\hbar \lbrack 1+i(\Delta _{c}-\delta
_{c}+B_{c}w_{0})]}\times \{-iw_{0}+\frac{2A^{2}\Omega _{c}^{2}w_{0}}{%
D(\delta _{c})}(1+i\Delta _{c})[1-i(\Delta _{c}+\delta _{c}+B_{c}w_{0})]
\nonumber \\
&&\times \lbrack 2-i(\delta _{c}+B_{c}w_{0})]\}.
\end{eqnarray}

The population inversion of exciton is determined by equation:

\begin{equation}
w_{0}=\frac{-2\Omega _{c}^{2}w_{0}}{1+(\Delta _{c}+B_{c}w_{0})^{2}}-1.
\end{equation}
The cubic equation has either a single root or three real roots. The latter
case just corresponds to the intrinsic optical bistability which arises from
the Coulomb interaction.

In terms of this model, we can also determine the light group velocity\cite%
{27,28}

\begin{equation}
v=\frac{c}{n+\omega _{s}(dn/d\omega _{s})},
\end{equation}%
where $n\approx 1+2\pi \chi $, and then
\begin{equation}
\frac{c}{v_{g}}=1+2\pi Re\chi (\omega _{s})_{\omega _{s}=\omega _{ex}}+2\pi
Re(\frac{d\chi }{d\omega _{s}})_{\omega _{s}=\omega _{ex}}.
\end{equation}

It clear from this expression for $v_{g}$ that when $Re\chi (\omega
_{s})_{\omega _{s}=\omega _{ex}}$ is zero and the dispersion is very steep
and positive, the group velocity is significantly reduced,and them
\begin{equation}
\frac{c}{v_{g}}-1=\frac{2\pi \omega _{ex}N\mu ^{2}T_{2}}{\hbar }Re(\frac{%
d\chi }{d\omega _{s}})_{\omega _{s}=\omega _{ex}}=\frac{\sum }{T_{2}}Re(%
\frac{d\chi }{d\omega _{s}})_{\omega _{s}=\omega _{ex}},
\end{equation}%
where
\[
\sum =\frac{2\pi \omega _{ex}N\mu ^{2}T_{2}^{2}}{\hbar }.
\]%
A weak pulse propagating through a distance $L$ with a reduced group
velocity is delayed as compared with propagation in free space by the time
delay $T_{g}$ which is often the quantity measured in experiments,
\begin{equation}
T_{g}=\frac{L}{c}(\frac{c}{v_{g}}-1)=\frac{2\pi \omega _{ex}NL\mu ^{2}T_{2}}{%
c\hbar }Re(\frac{d\chi }{d\omega _{S}})_{\omega _{s}=\omega _{ex}},
\end{equation}%
where $c$ is the free-space speed of light.

\section{Numerical Results and Discussions}

In the following, as an example, we consider a Au MNP with radius $a=7.5nm$
and the dielectric constants of the background medium and SQD: $\varepsilon
_{0}=1,$ and \ $\varepsilon _{s}=6$. For the relaxtion time and dipole
moment in SQD, we take $T_{2}=0.8ns$ and $\mu =er_{0}$, with $r_{0}=0.65nm$%
\cite{8}.

Figure 1 shows the effect of $B_{c}$ due to the distance between SQD and NMP
and the dielectric constants of the metal and the semiconductor quantum dot
on the linear absorption spectrum for three cases of the control field$%
(\Omega _{c}^{2}).$ It is clear from the figure that the absorption peak
move along to the larger $\Delta _{s}$$(\Delta _{s}=(\omega _{s}-\omega
_{ex})T_{2})$ with the larger value of $B_{c}.$ This implies that the SQD
and the NMP close more and the Coulomb interaction between them becomes
stronger. As a result, the Stark shift is significant.

The linear optical absorption$Im\chi ^{(1)}$ as a function of the detuning $%
\Delta _{s}$ for the case $\Delta _{c}=1.5,B_{c}=-0.1$ is presented in
Figure 2. In the figure the dotted line is a normal excitonic absorption in
the hybrid complex system as the control field is absent $(\Omega
_{c}^{2}=0).$ When we turn on the control field $(\Omega _{c}^{2}=5),$ a
non-absorption hole appears at $\Delta _{s}=0,$ thus the system becomes
transparent for the probe field. The hole is induced by coherent population
oscillation which is mediated by the plasmon-exciton interaction. The
physical reason is that the control field pump the electrons from the ground
state to the first excited state. Turning on the probe field, the probe
field will adjust the amplitude of the control field. The adjust function of
the probe field and the long dephasing time $T_{1}$ will make the electronic
population oscillating with the beat frequency $\delta $ between two energy
levels. This coherent population oscillation can lead to the decrease of the
absorption of the probe field. But here the plasmon effects will modify sucn
a CPO and is beneficial to the slow light (also see Fig.6 below). Figure 3
demonstrates the behavior of the real part of the linear optical
susceptibility $\chi ^{(1)}$ as a function of the detuning $\Delta _{s}$ for
$\Delta _{c}=1.5,B_{c}=-0.1,\Omega _{c}^{2}=5.$ At $\Delta _{s}=0,$ there is
a steep slope corresponding to a large dispersion, which results in slow
group velocity without absorption which shows in the solid line in Figure 2.

Figure 4 illustrates the imaginary part of $\chi ^{(1)}$ as a function of $%
\Delta _{s}$ with $\Omega _{c}^{2}=5,B_{c}=-0.1$for the three values of $%
\Delta _{c}$. Only $\Delta _{c}=1.5,$ the transparency will occur at $\Delta
_{s}=0$. This condition is important in the formation of the transparency at
the center of the exciton absorption, otherwise the transparency will
disappear at the center of exciton absorption. The imaginary part of the
linear optical susceptibility $\chi ^{(1)}$ as a function of the detuning $%
\Delta _{s}$ for the three values of $\Omega _{c}^{2}$ with $\Delta
_{c}=1.5,B_{c}=-0.1$ is shown in Figure 5(a). At the resonant condition $%
\Delta _{s}=0,$ the linear absorption decreases and the transparency window
of the absorption spectrum between two absorption peaks becomes broader with
larger value $\Omega _{c}^{2}.$Figure 5(b) shows the real part of the linear
optical susceptibility $\chi ^{(1)}$ as a function of the detuning $\Delta
_{s}$ for the three value of $\Omega _{c}^{2}$ with $\Delta
_{c}=1.5,B_{c}=-0.1$. At $\Delta _{s}=0,$ the slope of refractive index can
be flatter and the width between two peaks can be broader by increasing $%
\Omega _{c}^{2}$.

The group velocity index $n_{g}$ as a function of the Rabi frequency $\Omega
_{c}^{2}$ with $\Delta _{c}=1.5$ is plotted ( in units of $\sum $ ) for
three value of $B_{c}$ in the resonance case $(\Delta _{s}=0)$ in Figure 6.
It is shown that the group velocity is very sensitive to the Rabi frequency
of the control field and the exciton-plasmon interaction. The slope of the
dispersion become steeper as $\Omega _{c}^{2}$ decreases, leading to an
increasingly group velocity. The value of $B_{c}$ is smaller, which
corresponds to the interparticle interaction R is larger, so the local field
resulting in the Coulomb interaction is weaker. Because of the negative
value of $B_{c}$, this local field will make the total field weaken. So the
total field felt by the SQD is weaker with the larger absolute value of $%
B_{c}.$ That is the reason of smaller group velocity index with smaller
absolute value of $B_{c\text{.}}$

\section{Conclusions}

In conclusion, the slow light based on the coherent population oscillation
mediated by exciton-plasmon interaction in hybrid complex is investigated
theoretically. The numerical results show that the exciton-plasmon
interaction have a significant effect on the absorption spectrum and the
dispersion spectrum and slow light can be achieved in such a hybrid complex.
Finally, we hope that the proposed effect of this work will be testified in
the near future experiments.

\begin{acknowledgments}
The part of this work was supported by National Natural Science Foundation
of China(No.10774101) and the National Minister of Education Program for
Training Ph.D..
\end{acknowledgments}

\newpage \centerline{\large{\bf Figure Captions}}

Fig 1 The imaginary part of the linear optical susceptibility as a function
of the detuning $\Delta _{s}$ with parameters $\Delta _{c}=1.5,\Omega
_{c}^{2}=0$, for $B_{c}=0,-1,-7.$

Fig 2 The imaginary part of the linear optical susceptibility as a function
of the detuning $\Delta _{s}$ with parameters $\Delta _{c}=1.5,B_{c}=-0.1$,
for $\Omega _{c}^{2}=0,5.$

Fig 3 The real part of the linear optical susceptibility as a function of
the detuning $\Delta _{s}$ with parameters $\Delta _{c}=1,B_{c}=-0.1,\Omega
_{c}^{2}=5.$

Fig 4 The imaginary part of the linear optical susceptibility as a function
of the detuning $\Delta _{s}$ with parameters $\Omega _{c}^{2}=5,B_{c}=-0.1,$%
for $\Delta _{c}=1,1.5,2.$

Fig 5.a The imaginary part of the linear optical susceptibility as a
function of the detuning $\Delta _{s}$ with parameters $\Delta
_{c}=1,B_{c}=-0.1$, for $\Omega _{c}^{2}=6,7,8.$

Fig 5.b The real part of the linear optical susceptibility as a function of
the detuning $\Delta _{s}$ with parameters $\Delta _{c}=1,B_{c}=-0.1$, for $%
\Omega _{c}^{2}=6,7,8.$

Fig 6 The group velocity index $n_{g}(=c/\upsilon _{g})$ ( in units of $\sum
$ ) as a function of the detuning $\Delta _{s}$ with parameters$\Delta
_{c}=1.5,B_{c}=-0.1,-0.3,-0.5.$

\begin{figure}[tbp]
\includegraphics{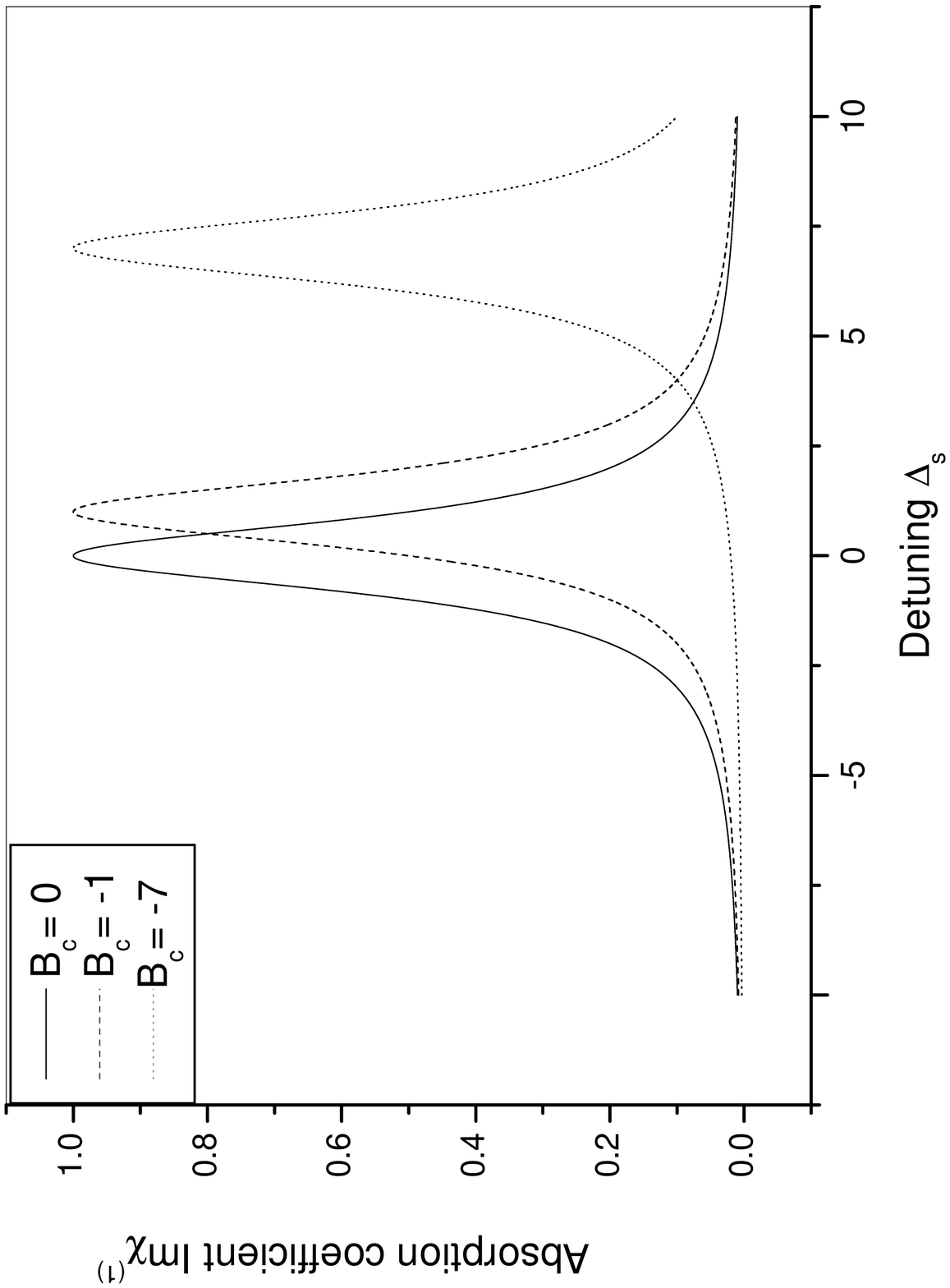}
\caption{}
\end{figure}

\clearpage
\begin{figure}[tbp]
\includegraphics{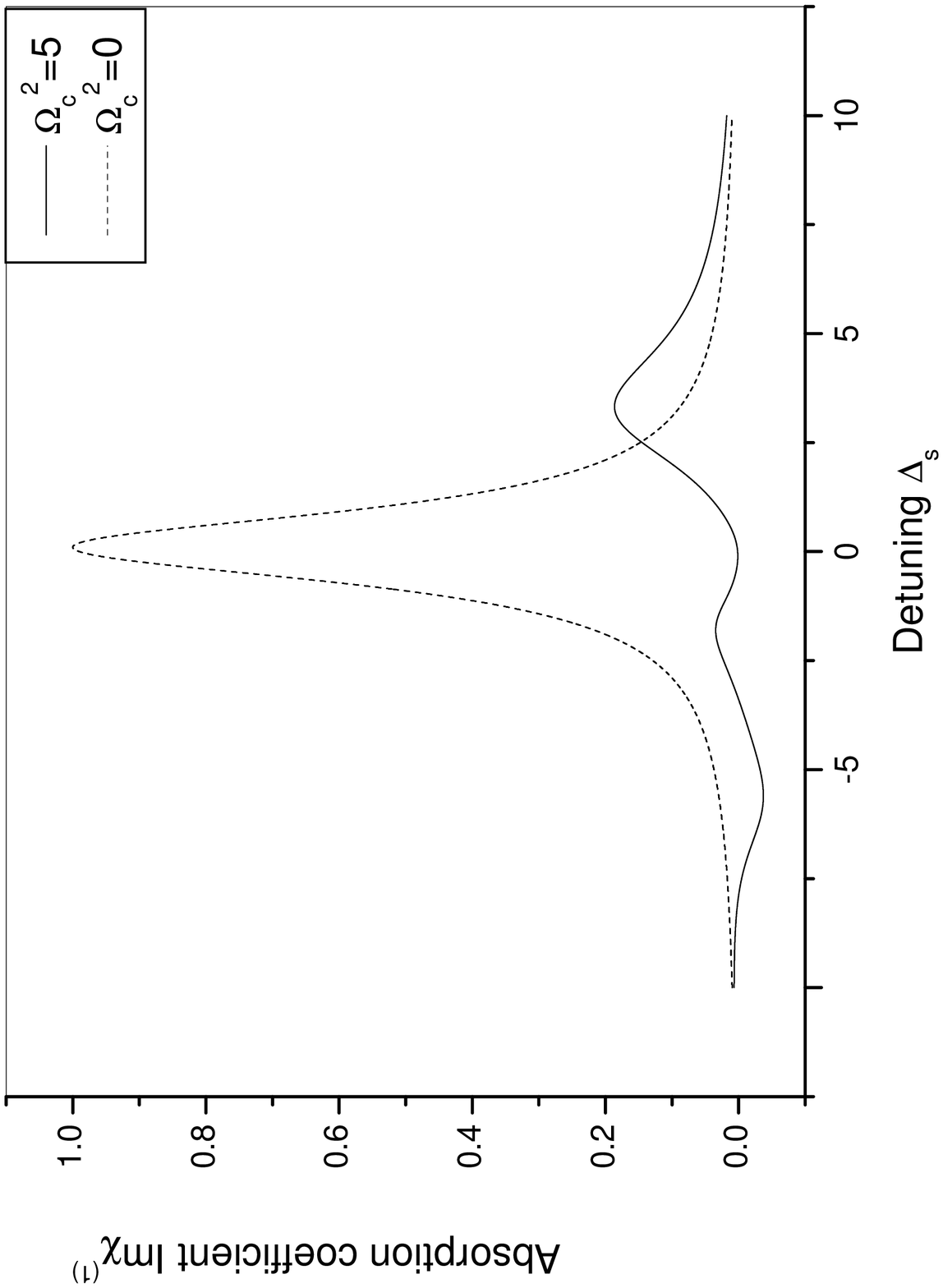}
\caption{}
\end{figure}

\clearpage
\begin{figure}[tbp]
\includegraphics{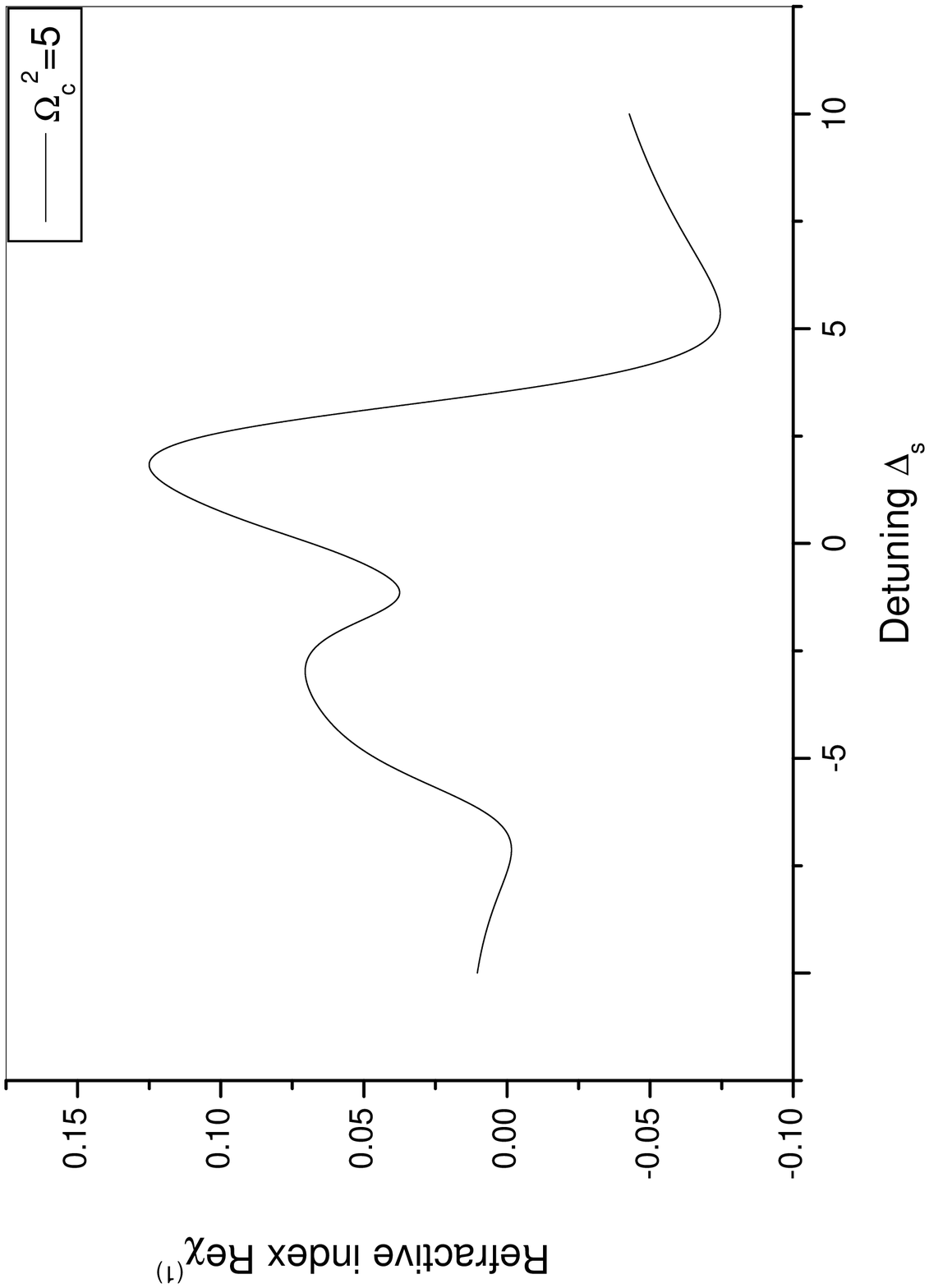}
\caption{}
\end{figure}

\clearpage
\begin{figure}[tbp]
\includegraphics{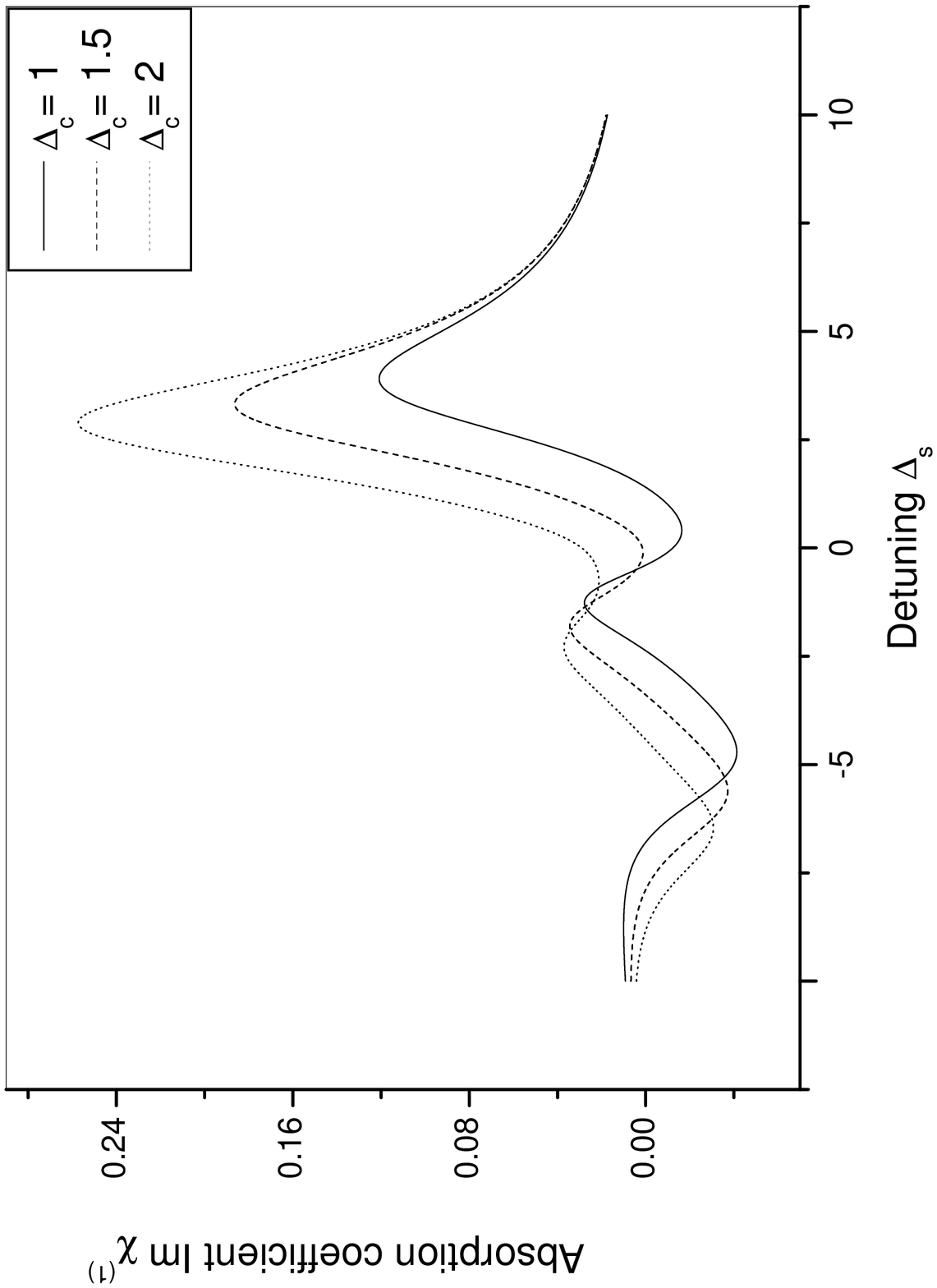}
\caption{}
\end{figure}

\clearpage
\begin{figure}[tbp]
\includegraphics{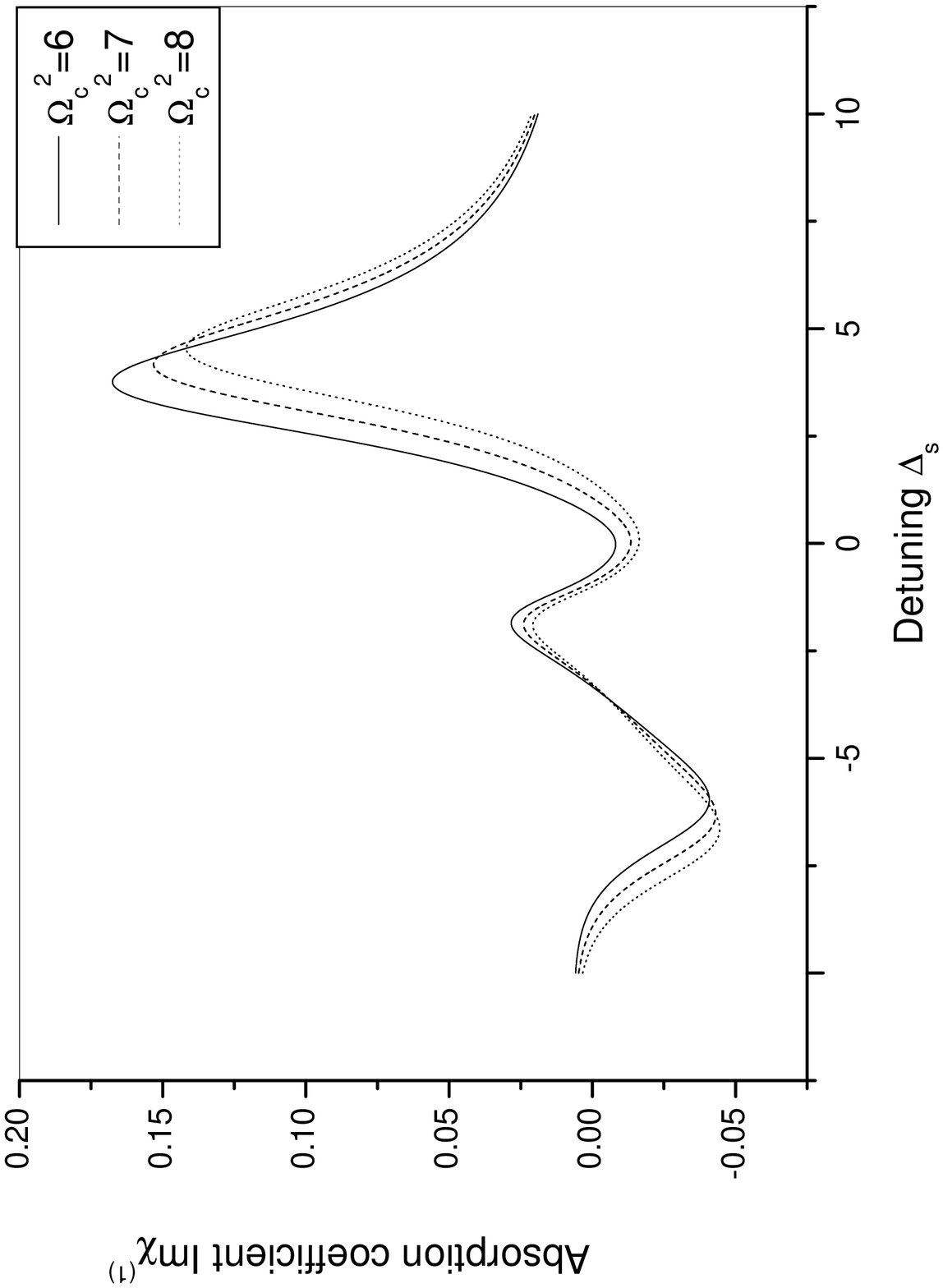}
\caption{}
\end{figure}

\clearpage
\begin{figure}[tbp]
\includegraphics{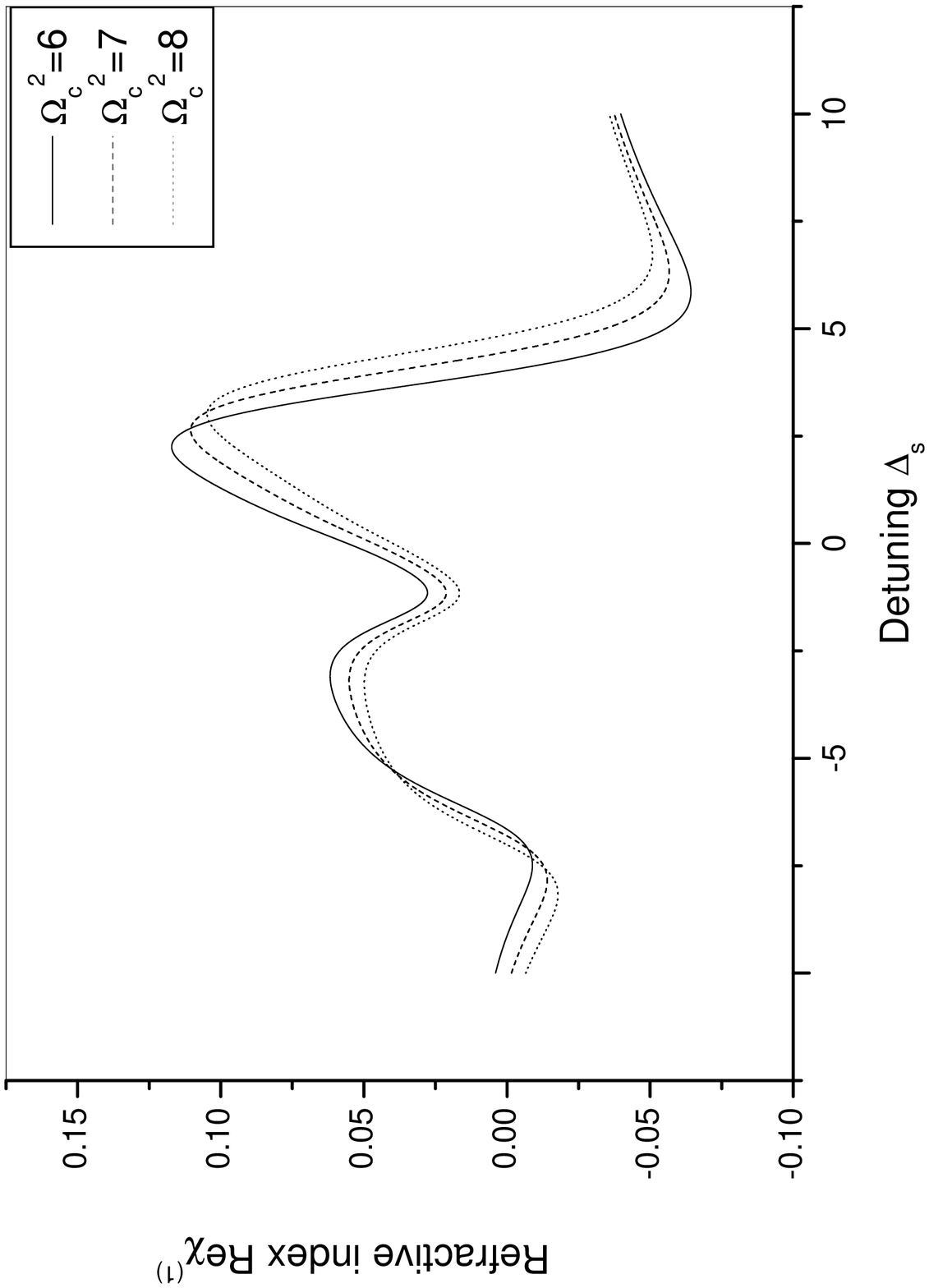}
\caption{}
\end{figure}

\newpage
\begin{figure*}[tbp]
\includegraphics{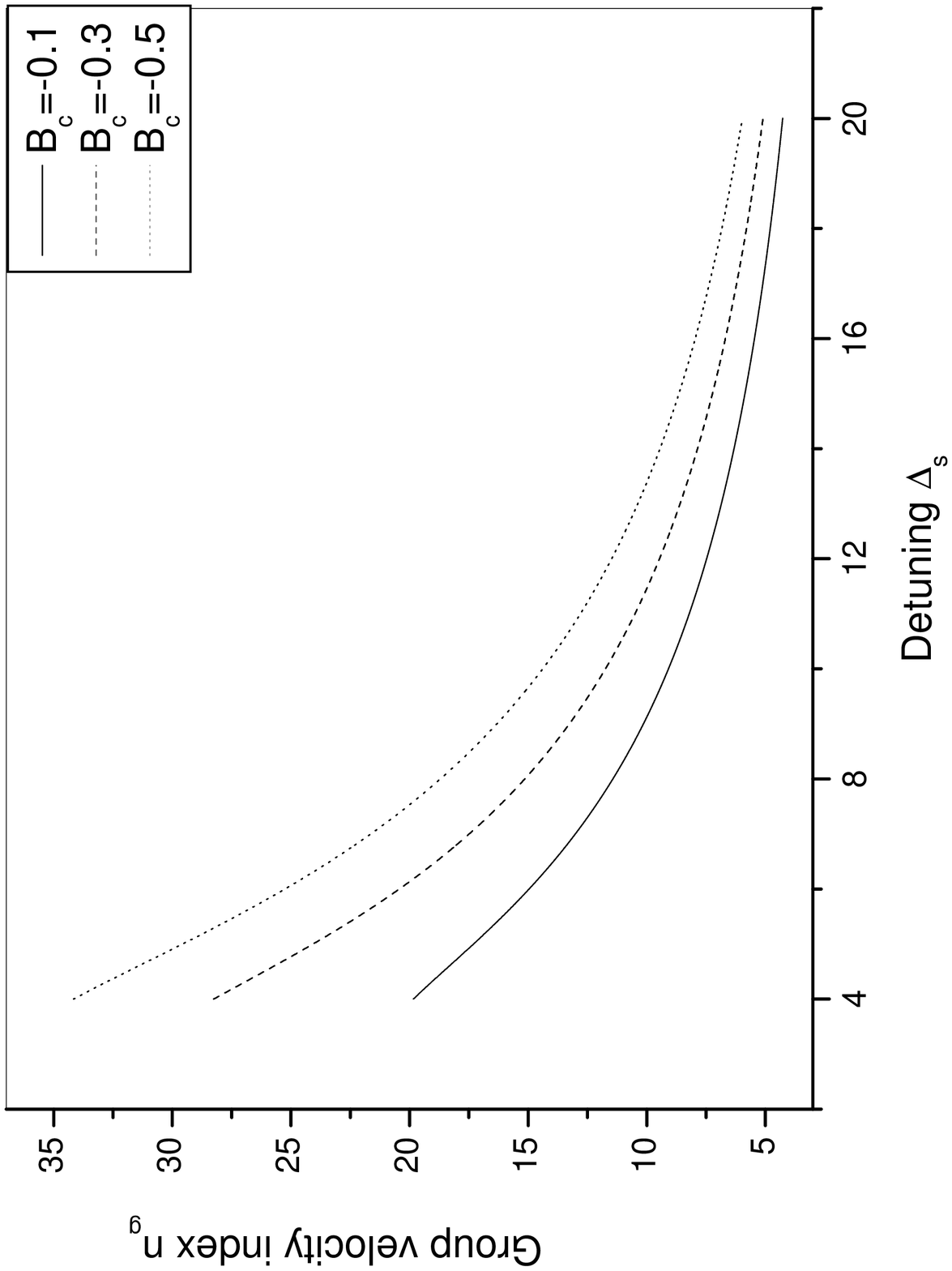}
\caption{}
\end{figure*}

\bibliographystyle{plain}
\bibliography{apssamp}
% Produces the bibliography via BibTeX.

\end{document}